\documentclass[runningheads]{llncs}

\usepackage[utf8]{inputenc}
\usepackage[T1]{fontenc}
\usepackage{graphicx}
\usepackage{amsmath,amssymb}
\usepackage{booktabs}
\usepackage{multirow}
\usepackage{xcolor}
\usepackage{url}

\begin{document}

\title{MindMelody: A Closed-Loop EEG-Driven System for Personalized Music Intervention}
\titlerunning{MindMelody}
\author{
Yimeng Zhang\textsuperscript{*},
Yueru Sun\textsuperscript{*},
Haoyu Gu\textsuperscript{*},
Zhanpeng Jin\textsuperscript{\textdagger}
\\
South China University of Technology, Guangzhou, China
\thanks{\textsuperscript{*}These authors contributed equally to this work.}
\thanks{\textsuperscript{\textdagger}Corresponding author.}
\thanks{Supported by National Training Program of Innovation and Entrepreneurship for Undergraduates (202510561174).}
}
\institute{}

\maketitle

\begin{abstract}
Driven by the escalating global burden of mental health conditions, music-based interventions have attracted significant attention as a non-invasive, cost-effective modality for emotion regulation and psychological stress relief. However, current digital music services rely on static preferences and fail to adapt to users' instantaneous psychological states. Furthermore, directly mapping electroencephalography (EEG) to music generation remains challenging due to severe paired-data scarcity and a lack of interpretability. To address these limitations, we propose MindMelody, a fully functional, closed-loop real-time system for EEG-driven personalized music intervention. MindMelody introduces an emotion-mediated semantic bridge. Specifically, a hybrid Transformer-GNN first decodes real-time EEG signals into global Valence-Arousal states and local temporal affect trajectories. These states are then fed into a Retrieval-Augmented Generation (RAG)-equipped Large Language Model (LLM) to formulate structured intervention plans. Subsequently, a novel Hierarchical EEG Controller injects global affect prefixes and local temporal guidance into a pretrained music backbone, enabling fine-grained controllable audio synthesis. Crucially, the system incorporates a continuous feedback loop that updates generation parameters on the fly based on the user's evolving EEG dynamics. Extensive experiments show that MindMelody improves control adherence and emotional alignment, and receives higher perceived helpfulness in a short-term listening setting, suggesting its promise as an adaptive affect-aware music generation framework.

\end{abstract}

\begin{figure}[t]
    \centering
    \includegraphics[width=0.85\textwidth]{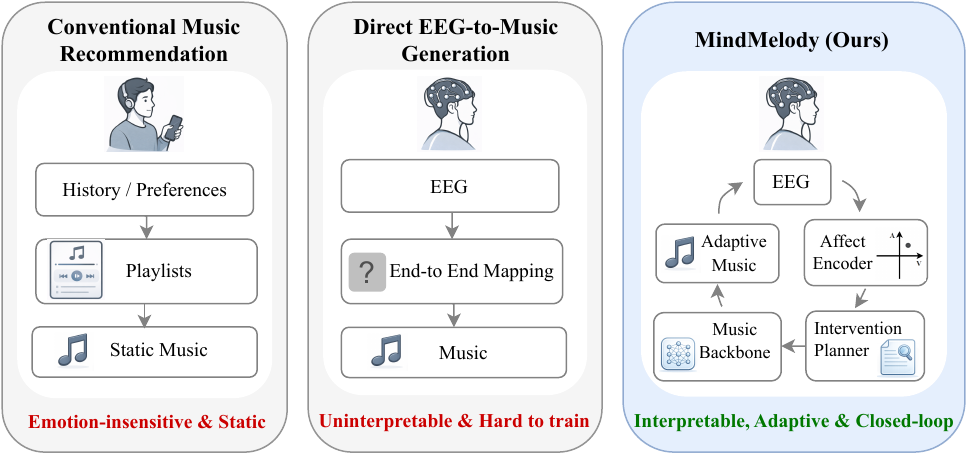}
    \caption{Comparison of three music intervention paradigms. Conventional recommendation is static and emotion-insensitive; direct EEG-to-music generation is difficult to interpret and train due to paired-data scarcity; MindMelody addresses these limitations through an emotion-mediated semantic bridge and a closed-loop adaptive intervention design.}
    \label{fig:paradigm_comparison}
\end{figure}

\section{Introduction}

\noindent Mental disorders remain a major public-health challenge worldwide. The World Health Organization reports that more than one billion people are living with mental health conditions, and common disorders such as anxiety and depression continue to impose substantial individual and societal burdens \cite{WHO2025}. In this context, music-based intervention has attracted sustained attention because it is non-invasive, low-cost, and suitable for repeated daily use. Prior psychophysiological studies have shown that music listening can influence the human stress response and facilitate autonomic recovery after acute stress exposure \cite{Thoma2013}. However, most current digital music services still rely on listening history, collaborative filtering, or coarse preference tags, rather than directly perceiving a user's instantaneous affective state. As a result, they are not well aligned with the dynamic demands of personalized and adaptive psychological intervention.

Electroencephalography (EEG) provides a promising pathway toward affect-aware intervention because it offers objective neural measurements with high temporal resolution and has become a core modality in affective computing and brain-computer interface research \cite{Koelstra2012,Zheng2015,Li2022}. Public benchmarks have enabled both dimensional and categorical emotion modeling from EEG. Nevertheless, robust cross-subject generalization and continuous affect estimation remain open challenges for practical deployment \cite{Lu2023}. In parallel, recent progress in retrieval-augmented language modeling and controllable music generation has made it increasingly feasible to transform affective states into semantically grounded music content \cite{Lewis2020,Elizalde2023,Copet2023,JASCO2024}. These developments motivate a critical research objective: converting EEG-derived affect into interpretable and musically meaningful control signals to formulate an automated intervention loop.

To address this, we propose and implement \textbf{MindMelody}, a fully functional, closed-loop real-time system for EEG-driven personalized music intervention. Rather than attempting an uninterpretable end-to-end mapping from EEG directly to waveforms—which suffers from severe paired-data scarcity—the proposed framework introduces an emotion-mediated semantic bridge via a novel hierarchical architecture. Specifically, the system utilizes a hybrid Transformer-GNN to decode real-time EEG signals into not only a global Valence-Arousal (VA) state but also a fine-grained local affect trajectory. The decoded states are subsequently fed into a Large Language Model (LLM) equipped with a Retrieval-Augmented Generation (RAG) mechanism, which searches a curated music therapy knowledge base to synthesize a structured intervention plan, such as tempo, dynamics, and instrumentation cues. Finally, rather than relying solely on generic text prompts, we introduce a Hierarchical EEG Controller that injects a global affect prefix and local temporal guidance into a pretrained controllable music backbone \cite{Copet2023,JASCO2024,Ding2024PETL}, enabling personalized music generation conditioned on both the structured intervention plan and EEG-derived affect dynamics.

Crucially, MindMelody operates as a dynamic closed-loop system rather than a static generator. It continuously monitors user EEG waveforms, tracks emotion trajectories in the VA coordinate system, and periodically collects subjective user feedback. This dynamic mechanism allows the system to adjust generation parameters on the fly, ensuring that the music intervention tightly adapts to the user's evolving psychological state. After the experience, the system automatically generates a comprehensive intervention report, offering quantitative emotional improvement metrics and personalized therapeutic suggestions.

Our contributions are summarized as follows:
\begin{itemize}
    \item We propose MindMelody, a comprehensive closed-loop system for EEG-guided digital music therapy that integrates real-time physiological decoding, knowledge-grounded structured planning, and hierarchical controllable audio synthesis.
    \item We introduce an emotion-mediated generative strategy that extracts both global affect and local temporal trajectories from EEG, utilizing a Hierarchical EEG Controller to bridge physiological dynamics and music evolution, effectively circumventing paired-data scarcity.
    \item We design a dynamic intervention mechanism with a real-time feedback loop, continuously updating intervention targets based on user EEG changes and subjective evaluations for precise and adaptive regulation.
\end{itemize}

\section{Related Work}

\subsection{EEG Decoding}

EEG-based emotion recognition has been widely studied in affective computing. Benchmark datasets such as DEAP enabled dimensional modeling of valence, arousal, dominance, and liking from physiological recordings collected during music-video stimulation \cite{Koelstra2012}, while SEED established a standard benchmark for categorical EEG emotion recognition \cite{Zheng2015}. Recent surveys show that EEG emotion modeling has evolved from handcrafted spectral and connectivity features toward end-to-end deep architectures, including convolutional, recurrent, graph-based, and Transformer-based models \cite{Li2022}. However, substantial inter-subject variability still leads to performance degradation when models are transferred to unseen users, making cross-subject learning and domain adaptation central to practical EEG-driven systems \cite{Lu2023,Ganin2016}. Moreover, intervention-oriented generation requires temporally structured affect representations rather than recognition accuracy alone.

\subsection{Affect-to-Plan}

Large language models have enabled more flexible generation of structured and semantically rich intervention content. Retrieval-Augmented Generation (RAG) is particularly relevant because it improves knowledge-intensive generation by integrating parametric language models with explicit external memory \cite{Lewis2020}. In the audio domain, CLAP aligns text and audio in a shared embedding space, making it possible to connect semantic descriptions with audio content through a unified representation \cite{Elizalde2023}. These advances suggest that affective states can be translated into music-oriented semantic controls through a language mediation layer. However, existing text mediation approaches rarely convert physiological affect estimates into structured intervention plans that are directly actionable for controllable music generation, such as tempo ranges, dynamic profiles, and instrumentation cues.

\subsection{Controllable Music}

Recent years have witnessed rapid progress in controllable music generation. MusicGen demonstrated that high-quality music can be generated directly from compressed discrete music tokens under text and melody conditions \cite{Copet2023}. JASCO further showed that global text descriptions can be combined with fine-grained local controls, substantially improving temporal controllability in text-to-music generation \cite{JASCO2024}. More recently, parameter-efficient transfer methods have been explored for music foundation models, suggesting that adapters, prompts, and related lightweight mechanisms can effectively adapt strong pretrained backbones to downstream tasks with reduced training cost \cite{Ding2024PETL}. These trends motivate our choice to place the main methodological novelty in an EEG-specific controller rather than in training a new music generator from scratch.

Alongside model advances, evaluation has also become increasingly important. Fr\'echet Audio Distance (FAD) is now widely used as a reference-free metric for generative audio quality, while human listening studies are commonly used to assess semantic alignment and perceptual realism \cite{Copet2023,JASCO2024,Kilgour2019}. A small but emerging line of work has explored direct music reconstruction from EEG using latent diffusion models \cite{Postolache2024}. In contrast, our work emphasizes a hierarchical route from EEG to affect, from affect to structured planning, and ultimately from planning to controllable music generation.

\section{Method}

\begin{figure}[t]
    \centering
    \includegraphics[width=0.9\textwidth]{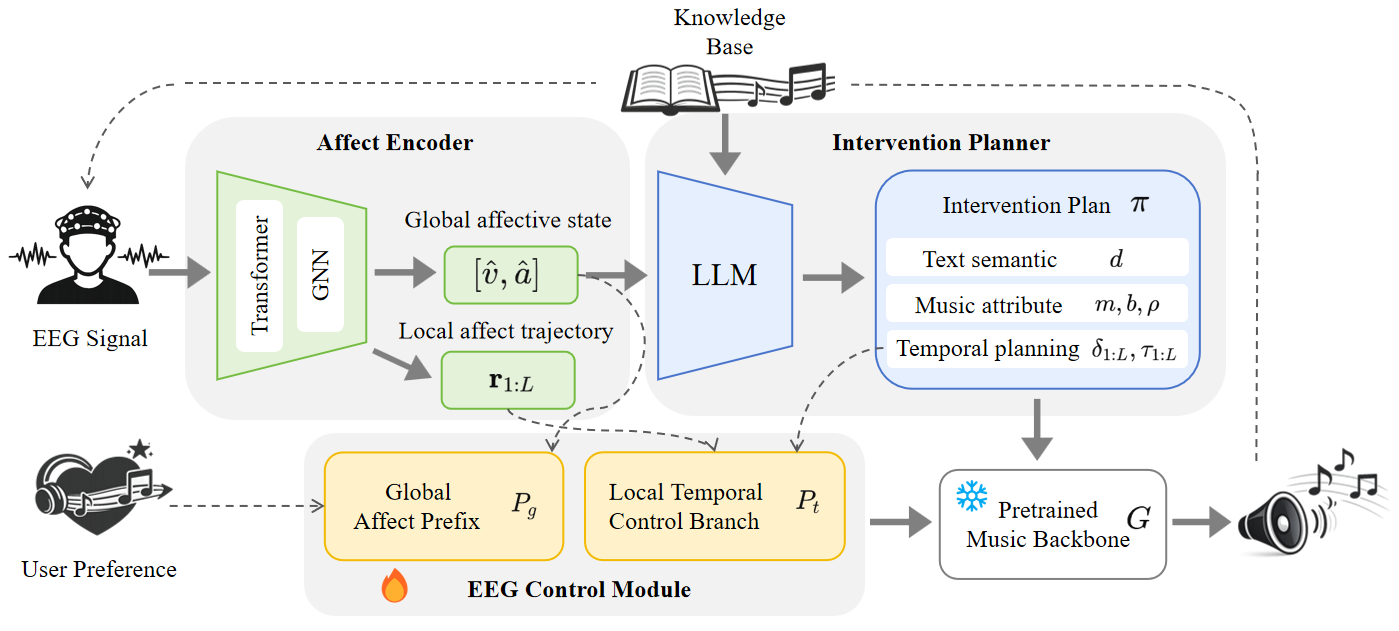}
    \caption{Overview of the proposed framework, consisting of an Affect Encoder, an Intervention Planner and an EEG Control Module.}
    \label{fig:overflow}
\end{figure}

\subsection{Overall Framework}

We formulate EEG-guided personalized music intervention as a hierarchical controllable generation problem. Given an EEG segment $X$, the goal is to generate a music sample $y$ that is consistent with the user’s current affective state and evolves toward the intended intervention target. Instead of directly mapping EEG to waveform or music tokens, we use affect as an intermediate semantic bridge. The system first decodes a global affective state and a local affect trajectory from EEG, then converts them into a structured intervention plan through a retrieval-augmented intervention planner, and finally injects these conditions into a pretrained music backbone through a hierarchical EEG controller.

The overall generation process is defined as
\begin{equation}
y = G(\pi, P_g, P_t),
\end{equation}
where $G$ denotes the pretrained music backbone, $\pi$ denotes the structured intervention plan, $P_g$ denotes the global affect prefix, and $P_t$ denotes the local temporal control sequence. This formulation explicitly models both the global affective direction of the music and its temporal evolution, which is well suited for psychological intervention scenarios.

\subsection{Affect Encoder}

Given a multichannel EEG segment $X\in\mathbb{R}^{C\times T}$, we first extract temporal features using a Transformer encoder and then model inter-channel relations with a graph neural network (GNN). The encoder outputs two levels of affective representation. The first is a global affective state,
\begin{equation}
\mathbf e_g=[\hat v,\hat a]\in\mathbb{R}^{2},
\end{equation}
which summarizes the current valence and arousal. The second is a local affect trajectory,
\begin{equation}
\mathbf r_{1:L}=\{\mathbf r_1,\mathbf r_2,\dots,\mathbf r_L\}, \quad \mathbf r_i\in\mathbb{R}^{2},
\end{equation}
which captures fine-grained temporal affect variation within the EEG segment. Since only trial-level VA annotations are available, the local trajectory supervision is constructed as weak temporal labels derived from window-level EEG dynamics under the consistency constraint of the global trial label.

\subsection{Intervention Planner}

A two-dimensional VA vector alone is insufficient to determine musically meaningful attributes such as instrument, tempo, texture density, and dynamics \cite{Melechovsky2024Mustango}. We therefore introduce a retrieval-augmented intervention planner that converts the predicted affect state into a structured intervention plan. The affect state is first reformulated into a semantic description, from which the system retrieves relevant snippets from a compact 1k-entry music-intervention knowledge base built from public music-therapy guidelines and affective music psychology literature \cite{Moore2015TFM,Liu2018TempoEmotion,DroitVolet2013MusicEmotion}. The retrieved context is combined with a fixed prompt and fed into Qwen2.5-7B-Instruct \cite{Qwen2_5_2024}, producing
\begin{equation}
\pi=\{d,m,b,\rho,\delta_{1:L},\tau_{1:L}\},
\end{equation}
where $d$ is the intervention description, $m$ denotes musical attributes, $b$ is the target tempo, $\rho$ is the texture density, $\delta_{1:L}$ is the section-wise dynamic plan, and $\tau_{1:L}$ is the target affect trajectory. Here, $\mathbf r_{1:L}$ describes the user's current short-term affect dynamics, whereas $\tau_{1:L}$ specifies the desired affect evolution for intervention.

\subsection{EEG Control Module}

Rather than training a music generator from scratch, we build on top of a pretrained music backbone and introduce a lightweight hierarchical EEG controller \cite{LiLiang2021,Hu2022LoRA}. The controller consists of a global affect prefix and a local temporal control branch. The global prefix is constructed as
\begin{equation}
P_g=M_g([\hat v,\hat a,p])\in\mathbb{R}^{K\times d},
\end{equation}
where $M_g$ is a lightweight projection module, $K$ is the number of virtual control tokens, and $d$ is the backbone hidden dimension. It controls the overall emotional direction and style. The local branch is constructed as
\begin{equation}
P_t=M_t([\mathbf r_{1:L},\tau_{1:L},\delta_{1:L}])\in\mathbb{R}^{L\times d},
\end{equation}
where $M_t$ is a lightweight temporal projection module. It is built from the EEG-derived trajectory, the planner-generated target trajectory, and the dynamic plan, and controls how the target emotion evolves over time. In practice, the global prefix is injected through projection-based conditioning, while the local branch modulates intermediate features via temporal conditioning and cross-attention \cite{LiLiang2021,Gheini2021}.

\subsection{Training Objective and Closed-Loop Update}

The framework is optimized with
\begin{equation}
\mathcal{L}
=\mathcal{L}_{gen}
+\lambda_1\mathcal{L}_{EEG}
+\lambda_2\mathcal{L}_{txt-aud}
+\lambda_3\mathcal{L}_{emo-align}
+\lambda_4\mathcal{L}_{ctrl},
\end{equation}
where the terms denote generation loss, EEG affect supervision, text-audio semantic consistency, affect alignment, and control adherence, respectively.

Training is performed in two stages. We first train the Transformer-GNN affect encoder with global VA supervision and weak local trajectory supervision, and then freeze it to optimize the EEG-conditioned generation modules on top of the pretrained backbone.

At inference time, EEG is first mapped to the global affect state and local trajectory, after which the planner generates a structured intervention plan and the backbone synthesizes music under joint conditioning. After listening, a new EEG segment is acquired and the post-listening affective state is estimated as $\mathbf e'_t=[\hat v'_t,\hat a'_t]$. Let $\mathbf e^\star$ denote the desired target state. We compute
\begin{equation}
\mathbf r_t=\mathbf e^\star-\mathbf e'_t,
\end{equation}
and update the next-round target by
\begin{equation}
\tilde{\mathbf e}_{t+1}=\mathbf e'_t+\alpha \mathbf r_t,
\end{equation}
where $\alpha\in(0,1]$ is the intervention step size. The updated global target is converted into the next-round plan, while the residual trend is mapped to the next local target trajectory, forming a closed loop of recognition, planning, generation, and feedback.

\section{Experiments}

\begin{table}[t]
\centering
\caption{Music generation quality and control-adherence results (mean $\pm$ std).}
\label{tab:gen_quality}
\vspace{2pt}
\footnotesize
\setlength{\tabcolsep}{4pt}
\resizebox{\textwidth}{!}{%
\begin{tabular}{lccccc}
\toprule
Method & FAD $\downarrow$ & CLAP-Sim $\uparrow$ & Emo-MSE $\downarrow$ & Dyn-Corr $\uparrow$ & Plan-Cons $\uparrow$ \\
\midrule
Text-only                   & $3.40 \pm 0.11$ & $0.320 \pm 0.016$ & $0.142 \pm 0.014$ & $0.41 \pm 0.07$ & $0.58 \pm 0.09$ \\
Text + static VA            & $3.33 \pm 0.07$ & $0.330 \pm 0.011$ & $0.121 \pm 0.012$ & $0.46 \pm 0.04$ & $0.63 \pm 0.05$ \\
Text + global affect prefix & $3.26 \pm 0.04$ & $0.340 \pm 0.009$ & $0.103 \pm 0.006$ & $0.52 \pm 0.04$ & $0.69 \pm 0.02$ \\
\textbf{Ours full}          & $\mathbf{3.18 \pm 0.05}$ & $\mathbf{0.350 \pm 0.008}$ & $\mathbf{0.082 \pm 0.005}$ & $\mathbf{0.63 \pm 0.02}$ & $\mathbf{0.78 \pm 0.04}$ \\
\bottomrule
\end{tabular}%
}

\end{table}

\subsection{Setup}

\subsubsection{Datasets}

We evaluate the proposed framework on both EEG affect data and music--text data. For EEG affect modeling, we train and evaluate the Affect Encoder on DEAP under a cross-subject setting. DEAP contains EEG recordings from 32 participants watching 40 one-minute music video excerpts, with valence and arousal for each trial~\cite{Koelstra2012}. For controllable music generation, we annotate 2,000 MusicCaps clips with valence--arousal scores on a 1--9 scale, following the DEAP rating protocol. Annotators were provided with written instructions and example anchors for low/high valence and low/high arousal. Clips were presented in randomized order, and the final label was obtained by averaging the three ratings. The resulting ICC(2,k) is 0.77, indicating good inter-rater agreement and supporting the reliability of the affect annotations. The annotated subset covers diverse genres and instrumentation patterns from MusicCaps, which helps reduce style-specific bias. This subset serves as auxiliary supervision for affect-aware semantic alignment and controllable generation, and is derived from the public MusicCaps corpus released with MusicLM~\cite{Agostinelli2023}.

\begin{table}[t]
\centering
\caption{Subjective evaluation and closed-loop intervention results (mean $\pm$ std).}
\label{tab:user_study}
\vspace{2pt}
\footnotesize
\setlength{\tabcolsep}{4pt}
\resizebox{\textwidth}{!}{%
\begin{tabular}{lccccc}
\toprule
Condition & Nat.-MOS $\uparrow$ & Emo.-MOS $\uparrow$ & Help. $\uparrow$ & $\Delta$Valence $\uparrow$ & Aro.-Dev. $\downarrow$ \\
\midrule
Human-selected playlist & $4.34 \pm 0.29$ & $4.06 \pm 0.33$ & $4.01 \pm 0.35$ & $0.18 \pm 0.07$ & $0.17 \pm 0.06$ \\
Text-only               & $3.89 \pm 0.37$ & $3.74 \pm 0.41$ & $3.68 \pm 0.44$ & $0.10 \pm 0.08$ & $0.27 \pm 0.09$ \\
Text + static VA        & $3.96 \pm 0.35$ & $3.88 \pm 0.39$ & $3.82 \pm 0.40$ & $0.14 \pm 0.08$ & $0.22 \pm 0.08$ \\
\textbf{Ours full}      & $\mathbf{4.12 \pm 0.31}$ & $\mathbf{4.21 \pm 0.28}$ & $\mathbf{4.18 \pm 0.30}$ & $\mathbf{0.22 \pm 0.06}$ & $\mathbf{0.14 \pm 0.05}$ \\
\bottomrule
\end{tabular}%
}

\end{table}

\subsubsection{Hyperparameters}

Each DEAP trial is segmented into 4-second windows with 2-second overlap. The EEG branch is trained with AdamW~\cite{Loshchilov2019} for 100 epochs using a learning rate of $1\times10^{-4}$, batch size 16, and weight decay $1\times10^{-4}$, with early stopping of 10 epochs. For music generation, we use MusicGen-medium(1.5B) \cite{Copet2023MusicGen} as the pretrained music backbone and optimize only the proposed EEG-specific hierarchical controller with AdamW~\cite{Loshchilov2019}, using a learning rate of $2\times10^{-4}$, batch size 8, and weight decay $1\times10^{-4}$. The generation length is fixed at 10 seconds, and the temporal EEG control sequence is resampled to the latent frame rate of the backbone. All models are trained on NVIDIA A100 GPUs.

\subsection{Metrics}
We report both objective and subjective metrics. FAD measures the distributional distance between generated and reference audio, where lower is better. CLAP-Sim denotes the cosine similarity between the intervention description $d$ and the generated audio $y$ in the CLAP embedding space, reflecting text-audio semantic consistency. To evaluate affect alignment, we use a frozen external music-affect estimator to predict the valence-arousal state $\hat{\mathbf e}(y)=[\hat v(y),\hat a(y)]$ from generated audio, and define
\begin{equation}
\mathrm{Emo\mbox{-}MSE}
=\frac{1}{2}\left[(\hat v(y)-v^\star)^2+(\hat a(y)-a^\star)^2\right],
\end{equation}
where $\mathbf e^\star=[v^\star,a^\star]$ is the target affective state. For temporal control, we segment each generated sample into $L$ sections and extract a section-wise loudness curve $\hat{\delta}_{1:L}(y)$ as a proxy for realized musical dynamics. We then compute
\begin{equation}
\mathrm{Dyn\mbox{-}Corr}
=\mathrm{PearsonCorr}\big(\delta_{1:L},\hat{\delta}_{1:L}(y)\big),
\end{equation}
where $\delta_{1:L}$ is the dynamic plan produced by the intervention planner. We further define Plan-Cons as a rule-based consistency score that measures whether the generated audio matches the planned tempo range, texture density, and section-wise dynamic trend. For subjective evaluation, Nat.-MOS, Emo.-MOS, and Help. denote 5-point Likert ratings of naturalness, emotion matching, and intervention helpfulness, respectively. In the closed-loop pilot study, $\Delta$Valence measures the post-intervention improvement in valence, and Aro.-Dev. denotes the absolute deviation between post-listening arousal and the target arousal.

\subsection{Main Results}
We evaluate the proposed system from three complementary perspectives: (1) EEG affect decoding, (2) automatic generation quality and control adherence, and (3) subjective quality together with closed-loop intervention effectiveness.

First, to verify whether the proposed Transformer-GNN affect encoder provides stable control signals, we conduct experiments on the DEAP dataset. The encoder achieves robust decoding performance, reaching $76.8 \pm 1.3\%$ valence accuracy and $72.4 \pm 1.5\%$ arousal accuracy, with concordance correlation coefficient (CCC)~\cite{Lin1989CCC} values of $0.43 \pm 0.02$ and $0.39 \pm 0.03$ for valence and arousal, respectively. These results suggest that the encoder captures affect-relevant EEG patterns and provides reliable global affect estimates together with temporally varying control cues for downstream planning and music generation.

Table~\ref{tab:gen_quality} summarizes the automatic evaluation results. While static VA conditioning already improves over the Text-only baseline, \textbf{Ours full} achieves the best overall performance with the hierarchical EEG controller. In particular, it reduces Emo-MSE from $0.142$ to $0.082$, while reaching the highest Dyn-Corr ($0.63$) and Plan-Cons ($0.78$). These results support the central claim of this work: the proposed system not only generates plausible music, reflected by a competitive FAD of $3.18$, but also substantially improves control adherence \cite{Melechovsky2024Mustango}. The gain mainly comes from explicitly modeling physiological temporal dynamics, which enables better semantic alignment and temporal controllability than static-VA baselines.

\begin{figure}[t]
   \centering
   \includegraphics[width=0.7\textwidth]{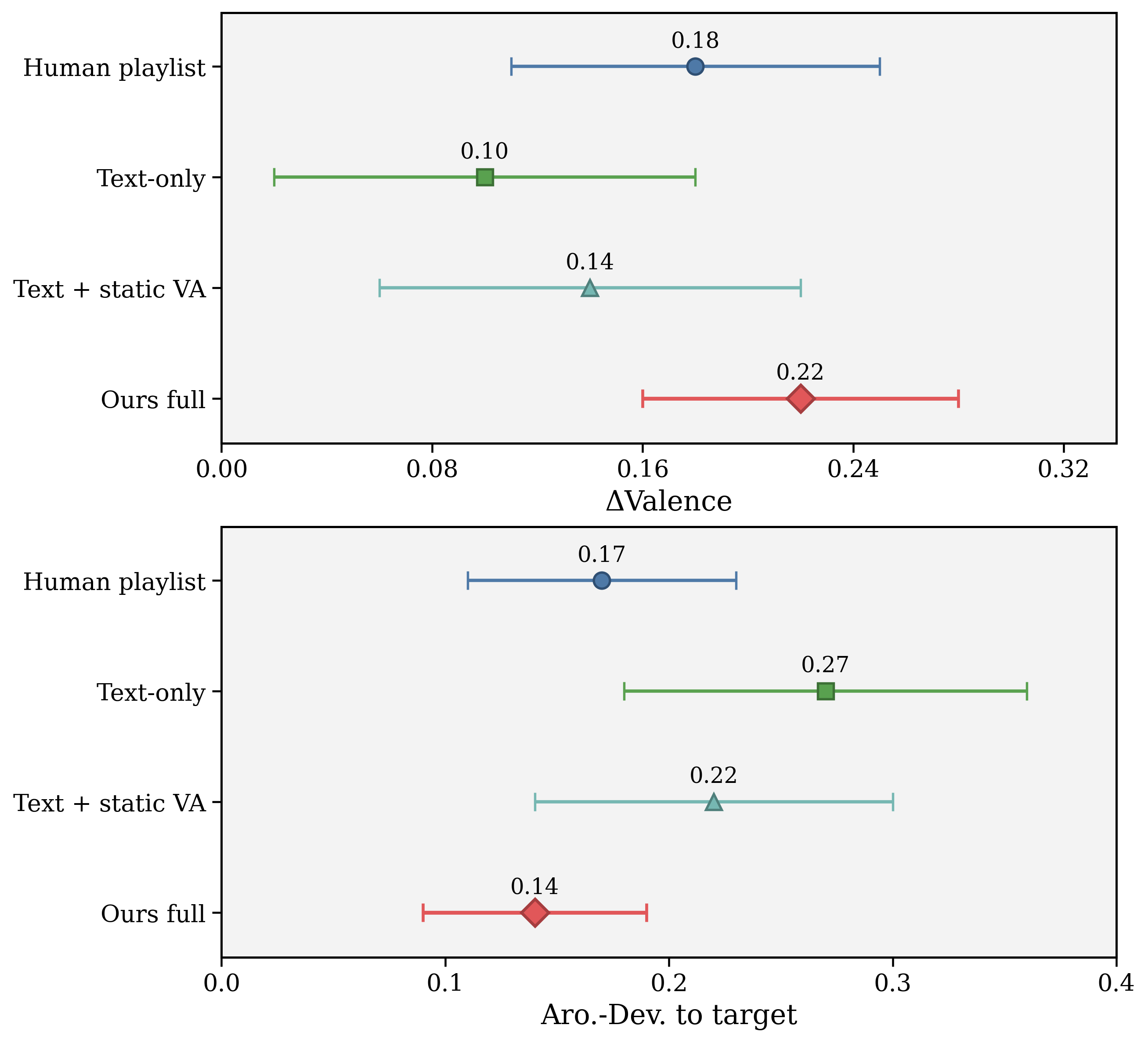}
   \caption{Closed-loop intervention results. Top: $\Delta$Valence (higher is better). Bottom: Aro.-Dev. (lower is better). Markers show mean values and error bars show standard deviations.}
   \label{fig:closed_loop_results}
\end{figure}

We conducted a pilot randomized within-subject user study with concealed system identities to evaluate subjective perception and short-term closed-loop affect regulation. Each participant experienced all four conditions: Human-selected playlist, Text-only, Text + static VA, and \textbf{Ours full}. In each trial, a baseline affective state was first estimated from EEG, after which the participant listened to a music intervention generated or selected under the corresponding condition. Post-listening affect was then re-estimated from EEG, and the participant provided subjective ratings. To mitigate order effects, the condition order was randomized on a per-participant basis. To reduce presentation bias, samples were displayed with anonymized condition labels and participants were not informed of the underlying system identity during evaluation. After each trial, participants rated naturalness (Nat.-MOS), emotion matching (Emo.-MOS), and perceived helpfulness (Help.) on 5-point Likert scales. We further computed $\Delta$Valence and Aro.-Dev. to assess short-term valence improvement and deviation from the target arousal state, respectively. As shown in Table~\ref{tab:user_study}, although the human-selected playlist achieves the highest Nat.-MOS ($4.34$), \textbf{Ours full} performs best in Emo.-MOS ($4.21$), Help. ($4.18$), $\Delta$Valence ($0.22$), and Aro.-Dev. ($0.14$), indicating its superior effectiveness for closed-loop affect regulation.

\subsection{Ablation Studies}

Ablation results in Table~\ref{tab:gen_quality} quantify the contribution of each key module. The temporal EEG control branch is the most critical for fine-grained controllability; removing it (see Text + global affect prefix) causes the sharpest drop in Dyn-Corr (from $0.63$ to $0.52$) and Plan-Cons (from $0.78$ to $0.69$). This confirms that explicitly modeling time-varying affect dynamics is more effective than relying on static prompts alone. The global affect prefix mainly supports overall emotional consistency, as removing it increases Emo-MSE from $0.103$ to $0.121$ and lowers Emo.-MOS. Finally, the retrieval-augmented intervention planner remains important for aligning physiological feedback with intervention intent, as it improves semantic coherence and plan interpretability. Overall, these results show that the advantage of the full model comes from the hierarchical synergy of physiological control, semantic planning, and affective conditioning, rather than from the pretrained backbone alone.

\section{Conclusion}

This paper presented MindMelody, a hierarchical closed-loop system for EEG-driven personalized music intervention. Rather than directly mapping EEG to audio, the proposed framework decomposes the task into affect decoding, structured intervention planning, and hierarchical music control, improving interpretability and tractability under limited paired data. Experimental results suggest that MindMelody improves control adherence, emotional alignment, and perceived helpfulness in short-term listening sessions, while showing encouraging potential for adaptive affect regulation. These findings should be interpreted as evidence from a non-clinical pilot study rather than as validation of therapeutic efficacy, and future work should include larger cohorts, longer-term evaluation, and clinically grounded protocols.

\end{document}